\pdfoutput=1
\documentclass[letterpaper, 10 pt, conference]{ieeeconf}  

\IEEEoverridecommandlockouts                              

\usepackage{graphicx}
\usepackage{url}
\usepackage{times}
\usepackage{epstopdf}
\usepackage{algorithmic}
\usepackage{amsmath} 
\usepackage{amssymb}  
\usepackage{balance}

\newtheorem{theorem}{Theorem}

\newtheorem{definition}[theorem]{Definition}

\newcommand{\eat}[1]{}

\makeatletter
\def\@begintheorem#1#2{\sl \trivlist \item[\hskip \labelsep{\bf #1\ #2:}]}
\def\@opargbegintheorem#1#2#3{\sl \trivlist
      \item[\hskip \labelsep{\bf #1\ #2\ #3:}]}
\makeatother

\title{\LARGE \bf
Optimizing Learned Bloom Filters by Sandwiching
}

\author{Michael Mitzenmacher{$^1$}\thanks{{$^1$}School~of Engineering and Applied Sciences, Harvard University.  
Supported in part by NSF grants CCF-1563710, CCF-1535795, and CCF-1320231.
This work was done while visiting Microsoft Research.}}

\begin{document}

\maketitle
\thispagestyle{empty}
\pagestyle{empty}

\begin{abstract}
We provide a simple method for improving the performance of
the recently introduced {\em learned Bloom filters}, by 
showing that they perform better when the learned function
is sandwiched between two Bloom filters.
\end{abstract}

\section{Introduction}

Recent work has introduced the {\em learned Bloom
filter} \cite{TCFLIS}.  As formalized in \cite{mypaper}, the learned
Bloom filter, like a standard Bloom filter, provides a compressed
representation of a set of keys ${\cal K}$ that allows membership
queries.  Given a key $y$, a learned Bloom filter will always
return yes if $y$ is in ${\cal K}$, and will generally return
no if $y$ is not in ${\cal K}$, but may return false positives.  
What makes a learned Bloom filter interesting is that it uses a
function that can be obtained by ``learning'' the set ${\cal K}$ to
help determine the appropriate answer.  Specifically, we recall
the following definition from \cite{mypaper}:

\begin{definition}
\label{def:lbf}
A {\em learned Bloom filter} on a set of positive keys ${\cal K}$ and
negative keys ${\cal U}$ is a function $f:U \rightarrow [0,1]$ and
threshold $\tau$, where $U$ is the universe possible query keys, and
an associated standard Bloom filter $B$, referred to as a backup filter.  The 
backup filter is set to hold the set of keys $\{z: z \in {\cal K}, f(z) < \tau\}$.  
For a query $y$, the learned Bloom filter returns that $y \in {\cal K}$
if $f(y) \geq \tau$, or if $f(y) < \tau$ and the backup filter
returns that $y \in {\cal K}$.  The learned Bloom filter returns 
$y \notin {\cal K}$ otherwise.
\end{definition}

\smallskip

In less mathematical terms, a learned Bloom filter consists of
pre-filter before a Bloom filter, where the backup Bloom filter now
acts to prevent false negatives.  The pre-filter suggested
in \cite{TCFLIS} comes from a neural network and estimates the
probability an element is in the set, allowing the use of a backup
Bloom filter that can be substantially smaller than a standard Bloom
filter for the set of keys ${\cal K}$.  If the function $f$ has a
sufficiently small representation, then the learned Bloom filter can
be correspondingly smaller than a standard Bloom filter for the same set.

\begin{figure*}[t]
        \centering
        \includegraphics[width=0.9\textwidth]{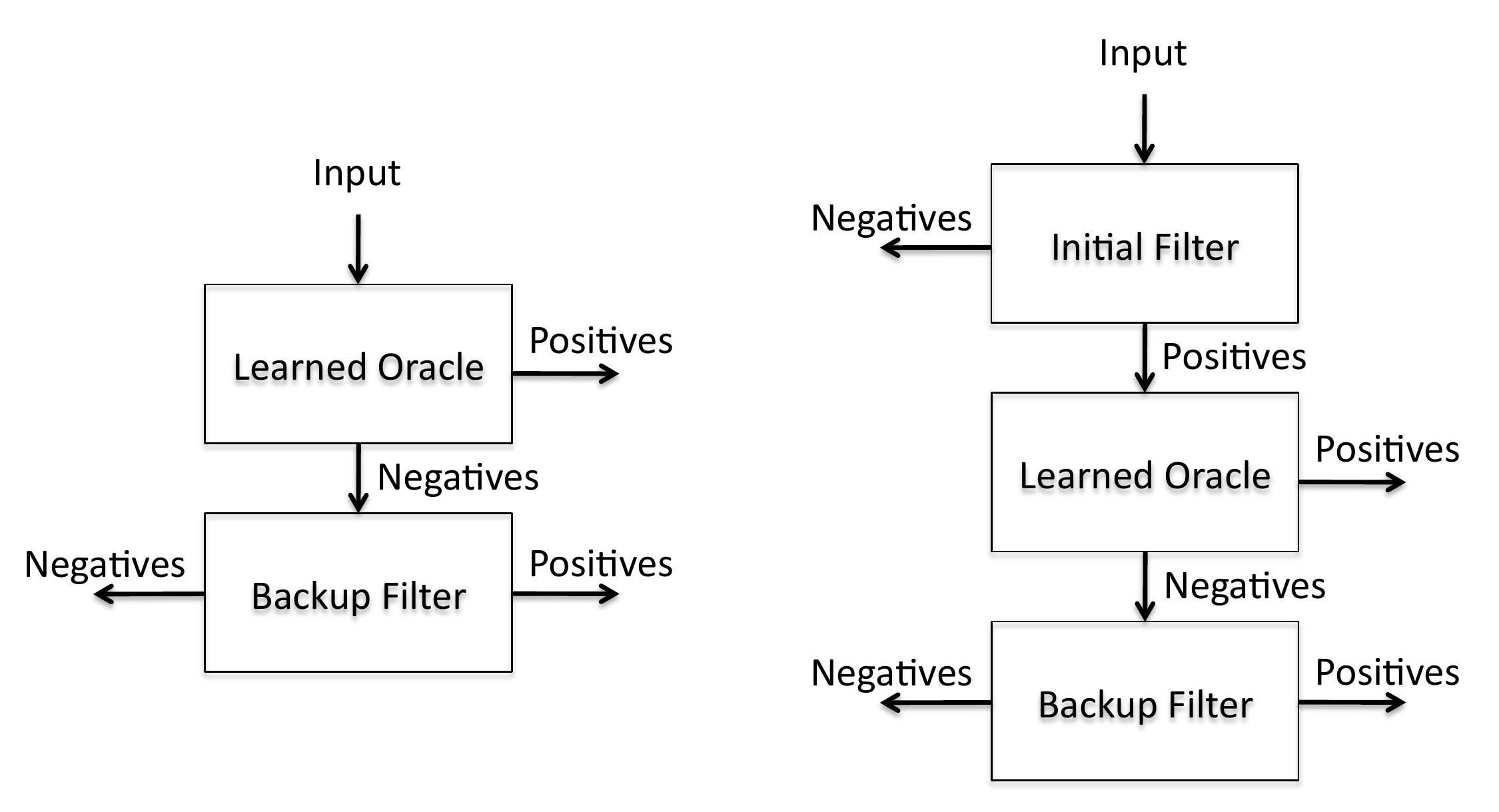}
                \caption{The left side shows the original learned Bloom filter.  The right side shows the sandwiched learned Bloom filter.}
                \label{fig:diagram}
\end{figure*}

Given this formalization of the learned Bloom filter, and the
additional analysis from \cite{mypaper} for determining the false
positive rate of a learned Bloom filter, it seems natural to ask
whether this structure can be improved.  Here we show, perhaps
surprisingly, that a better structure is to use a Bloom filter before
using the function $f$, in order to remove most queries for keys not
in ${\cal K}$.  We emphasize that this {\em initial Bloom filter} does
not declare that an input $y$ is in ${\cal K}$, but passes on all
matching elements the learned function $f$, and it returns $y \notin
{\cal K}$ when the Bloom filter shows the element is not in ${\cal
K}$.  Then, as before, we use the function $f$ to attempt to
remove false positives from the initial Bloom filter, and then use the
backup filter to allow back in keys from ${\cal K}$ that were false
negatives for $f$.  Because we have two layers of Bloom filters
surrounding the learned function $f$, we refer to this as a {\em
sandwiched learned Bloom filter}.  The sandwiched learned Bloom filter
is represented pictorially in Figure~\ref{fig:diagram}.

In hindsight, our result that sandwiching improves performance makes
sense.  The purpose of the backup Bloom filter is to remove the false
negatives arising from the learned function.  If we can arrange to
remove more false positives up front, then the backup Bloom filter can
be quite porous, allowing most everything that reaches it through, and
therefore can be quite small.  Indeed, our analysis shows that the
backup filter can be remarkably small, so that as the budget of bits
available for the Bloom filters increases, any additional bits should
go to the initial Bloom filter.  We present our analysis below.

\section{Analyzing Sandwiched Learned Bloom Filters}

We model the sandwiched learned Bloom filter as follows.  The middle
of the learned Bloom filter we treat as an {\em oracle} for the keys
${\cal K}$, where $|{\cal K}|=m$. For keys not in ${\cal K}$ there is
an associated false positive probability $F_p$, and there are $F_n m$
false negatives for keys in ${\cal K}$.  (The value $F_n$ is like a
false negative probability, but given ${\cal K}$ this fraction is
determined and known according to the oracle outcomes.)  This oracle
can represent the function $f$ associated with
Definition~\ref{def:lbf} for learned Bloom filters, but might also
represent other sorts of filter structures as well.  Also, as
described in \cite{mypaper}, we note that in the context of a learned
Bloom filter, the false positive rate is
necessarily tied to the query stream, and so in general may be an
empirically determined quantity; see \cite{mypaper} for further
details and discussion on this point.  
Here we show how to optimize over a single oracle,
although in practice we may possibly
choose from oracles with different values $F_p$ and $F_n$, in
which case we can optimize for each set of value and choose
the best suited to the application.  

We assume a total budget of $bm$ bits to be divided between an initial
Bloom filter of $b_1m$ bits and a backup Bloom filter of $b_2m$ bits.
To model the false positive rate of a Bloom filter that uses $j$ bits
per stored key, we assume the false positive rate falls as $\alpha^j$.  This
is the case for a standard Bloom filter (where $\alpha \approx 0.6185$
when using the optimal number of hash functions, as described in the
survey \cite{BroderMitzenmacher}), as well as for a static Bloom filter built using a
perfect hash function (where $\alpha = 1/2$, again described
in \cite{BroderMitzenmacher}).  The analysis can be modified to handle other functions
for false positives in terms of $j$ in a straightforward manner.  It
is important to note that if $|{\cal K}| = m$, the backup Bloom filter
only needs to hold $mF_n$ keys, and hence we take the number
of bits per stored key to be $b_2/F_n$.  Note that if we find the best
value of $b_2$ is $b$, then no initial Bloom filter is needed, but otherwise,
an initial Bloom filter is helpful.  

The false positive rate of a sandwiched learned Bloom filter is then 
$$\alpha^{b_1}(F_p + (1-F_p)\alpha^{b_2/F_n}).$$
To see this, note that for $y \notin {\cal K}$, $y$ first has to 
pass through the initial Bloom filter, which occurs with probability 
$\alpha^{b_1}$.  Then $y$ either causes a false positive from the
learned function $f$ with probability $F_p$, or with remaining
probability $(1-F_p)$ it yields a false positive on the backup
Bloom filter, with probability $\alpha^{b_2/F_n}$.

As $\alpha, F_p, F_n$ and $b$ are all constants for the purpose of this analysis,
we may optimize for $b_1$ in the equivalent expression
$$F_p \alpha^{b_1} + (1-F_p)\alpha^{b/F_n}\alpha^{b_1(1-1/F_n)}.$$
The derivative with respect to $b_1$ is 
$$F_p (\ln \alpha) \alpha^{b_1} + (1-F_p)\left (1- \frac{1}{F_n} \right)\alpha^{b/F_n}(\ln \alpha)\alpha^{b_1(1-1/F_n)}.$$
This equals 0 when 
$$\frac{F_p}{(1-F_p)\left (\frac{1}{F_n} -1 \right)} = \alpha^{(b-b_1)/F_n} = \alpha^{b_2/F_n}.$$
This yields that the false positive rate is minimized when
$$b_2 = F_n \log_\alpha \frac{F_p}{(1-F_p)\left (\frac{1}{F_n} -1 \right)}.$$

This result may be somewhat surprising, as here we see that $b_2$ is a
constant, independent of $b$.  That is, the number of bits used for
the backup filter is not a constant fraction of the total budgeted
number of bits $bm$, but a fixed number of bits; if the number of
budgeted bits increases, one should simply increase the size of the
initial Bloom filter as long as the backup filter is appropriately
sized.

In hindsight, returning to the expression for the false positive rate
$$\alpha^{b_1}(F_p + (1-F_p)\alpha^{b_2/F_n}),$$ we can see the
intuition for why this would be the case.  If we think of sequentially
distributing the $bm$ bits among the two Bloom filters, the expression
shows that bits assigned to the initial filter (the $b_1$ bits) reduce
false positives arising from the learned function (the $F_p$ term) 
as well as false positives arising subsequent to the learned function
(the $(1-F_p)$ term), while the backup filter only reduces false
positives arising subsequent to the learned function.
Initially we would provide bits to the backup filter to reduce the $(1-F_p)$ rate
of false positives subsequent to the learned function.  Indeed, 
bits in the backup filter drive down this
$(1-F_p)$ term rapidly, because the backup filter
holds fewer keys from the original set, leading to the $b_2/F_n$ (instead
of just a $b_2$) in the exponent in the expression $\alpha^{b_2/F_n}$.
Once the false positives coming through 
the backup Bloom filter reaches an appropriate level, which, by
plugging in the determined optimal value for $b_2$, we find is $$F_p
/\left(\frac{1}{F_n} -1 \right),$$ then the tradeoff changes.  At that point
the gains from reducing the false positives from the backup Bloom filter
are smaller than the gains obtained by using the initial Bloom filter.

As an example, using numbers roughly corresponding to settings tested
in \cite{TCFLIS}, suppose we have a learned function $f$ where $F_n =
1/2$ and $F_p = 1/100$.  For convenience we consider $\alpha = 1/2$
(that corresponds to perfect hash function based Bloom filters).
Then $$b_2 = (\log_2 99) /2 \approx 3.315.$$ Depending on our Bloom
filter budget parameter $b$, we obtain different levels of performance
improvement by using the initial Bloom filter.  At $b = 8$ bits per
key, the false positive rate drops from approximately $0.010015$ to
$0.000777$, over an order of magnitude.  Even at $b= 6$ bits per key,
the false positive rate drops from approximately $0.010242$ to
$0.003109$.

If one wants to consider a fixed false positive rate and consider the
space savings from using the sandwiched approach, that is somewhat
more difficult.  The primary determinant of the overall false positive
rate is the oracle's false positive probability $F_p$.  The sandwich
optimization allows one to achieve better overall false positive with
a larger $F_p$; that is, it can allow for weaker, and correspondingly
smaller, oracles.  

A possible further advantage of the sandwich approach is that it makes
learned Bloom filters more robust. As discussed in \cite{mypaper}, if
the queries given to a learned Bloom filter do not come from the same
distribution as the queries from the test set used to estimate the
learned Bloom filter's false positive probability, the actual false
positive probability may be substantially larger than expected.  The
use of an initial Bloom filter mitigates this problem, as this issue
affects the smaller number of keys that pass the initial Bloom filter.

In any case, we suggest that given that the sandwich learned Bloom
filter is a relatively simple modification if one chooses to use a
learned Bloom filter, we believe that the sandwiching method will
allow greater application of the learned Bloom filter methodology.

\end{document}